\documentclass{jps-cp}
\usepackage{color}

\title{Kondo Screening of Local Moments in a Triangular Triple Quantum Dot Connected to Normal and Superconducting Leads}

\author{Masashi \textsc{Hashimoto}$^{1}$, Yoshimichi \textsc{Teratani}$^{1,2}$, Masaya \textsc{Shirotani}$^{1}$, Yukihiro \textsc{Nakata}$^{1}$, Masashi \textsc{Shimamoto}$^1$, Yoichi \textsc{Tanaka}$^{3}$, Yasuhiro \textsc{Yamada}$^{4}$ and Akira \textsc{Oguri}$^{1,2}$}

\inst{
$^{1}$Department of Physics, Osaka City University, Osaka, 558-8585, Japan\\
$^{2}$
NITEP, Osaka Metropolitan University,  Osaka, 558-8585, Japan \\
$^{3}$Advanced Simulation Technology of Mechanics R$\&$D, Co. Ltd.,  Tokyo, 112-0002, 
Japan\\
$^{4}$NTT Basic Research Laboratories, NTT Corporation, Atsugi, Kanagawa, 243-0198, 
Japan}

\recdate{ July 30, 2022}

\abst{
We study the interplay between the Kondo and superconducting (SC) proximity effects, 
taking place in a triangular triple quantum dot (TTQD) connected  
to one normal and two  SC leads.  This system shows various quantum phases. 
Without the SC leads, 
the lowest two states that belong to the different spin sectors, 
$S=0$ and $S=1/2$,  
become energetically very close to each other  near half filling. 
The singlet one is a Kondo-screened state 
by conduction electrons from the normal lead, 
and the doublet one is a resonating valence bond state 
with unpaired free spin which 
remains unscreened. 
Furthermore, when one additional electron enters the TTQD,  
the ground state becomes a doublet  in which 
the $S=1$ local moment due to the Nagaoka ferromagnetism 
is partially screened by conduction electrons.
The Cooper pairs penetrating into the TTQD from the SC leads 
reconstruct the wavefunctions and vary the phase boundaries between 
these quantum states in the parameter space.
We calculate  ground-state phase diagrams using the numerical renormalization group, 
and show that the SC proximity effect induces a reentrant transition 
 in-between the  three- and four-electron fillings.

}

\kword{Quantum dot, Kondo effect, Superconductivity, 
Nagaoka ferromagnetism}

\begin{document}
\maketitle

\section{Introduction}

Triangular triple quantum dot (TTQD)  is a system 
that  has been studied over a decade 
theoretically\cite{Kuzmenko2006,Zitko2008,Mitchell2009,Numata2009,Oguri2011,Koga2013,Mitchell2013,Koga2016}
and experimentally \cite{Amaha2008, Seo2013, Kotzian2016}. 
The TTQD  features rich internal degrees of freedoms, 
such as an orbital along the triangular loop, and 
spin and charge degrees of freedoms 
including the total occupation number $N_d$ ($= 0, 1, . . . , 6$), 
 which can induce various kinds of Kondo effects. 

At half-filling $N_d =3$,   
the ground state of the isolated TTQD cluster  
for which all the leads are disconnected 
has a four-fold degeneracy caused by 
the  $C_{3v}$ symmetry of an equilateral triangle. 
This symmetry is broken when a single normal lead is connected to one 
of the dots in the TTQD,  
and thus the four states are separated into two spin states.  
One is the singlet bond (SB) state 
with an unpaired spin, 
which is eventually screened by conduction electrons,
in the dot connected to the normal lead. 
The other is the resonating valence bond (RVB) state 
with unpaired spin 
that  remains unscreened
 in the dots away from the normal lead.   
Furthermore, the Kondo effect of  an $S = 1$ high-spin Nagaoka state 
 occurs at $N_d \simeq 4$ \cite{Numata2009,Oguri2011}.

The Cooper pairs 
that can penetrate 
into the TTQD from 
superconducting (SC) leads give further variety  
 to the low-energy states. 
The SC proximity effects 
in triple-dot systems have theoretically been examined 
in some special situations so far 
\cite{Oguri2015,Yi2013, Bocian2018,Wrzesniewski2022}. 
However, the interplay between the Kondo and the SC proximity effects 
have not been fully understood.  
In this report, we clarify how the ground state
 evolves as a result 
of the interplay between the Kondo and SC proximity effects.

\begin{figure}[t]
	\centering
	\begin{tabular}{ccc}
		\includegraphics[width=0.23\linewidth]{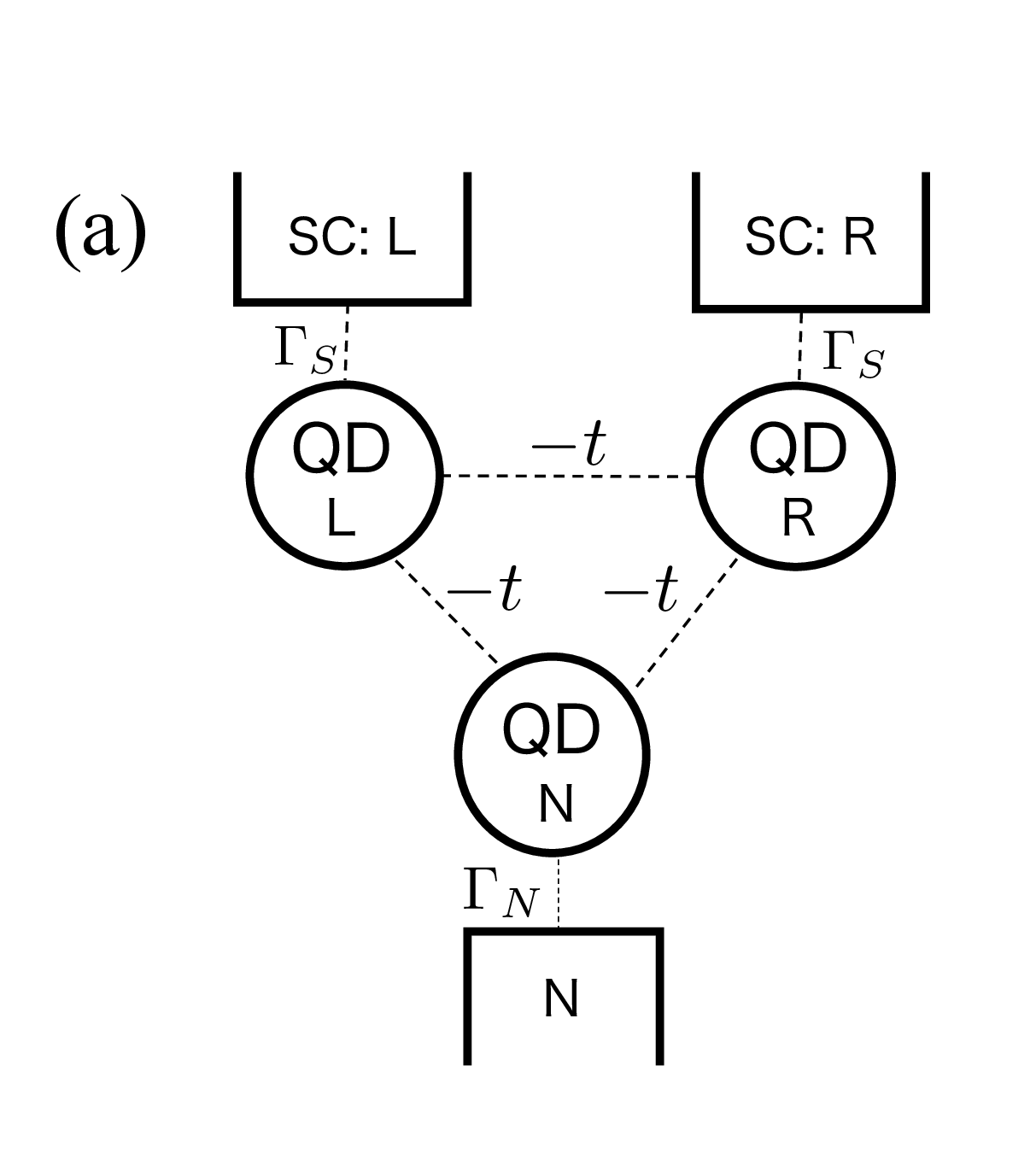}&
		\includegraphics[width=0.32\linewidth]{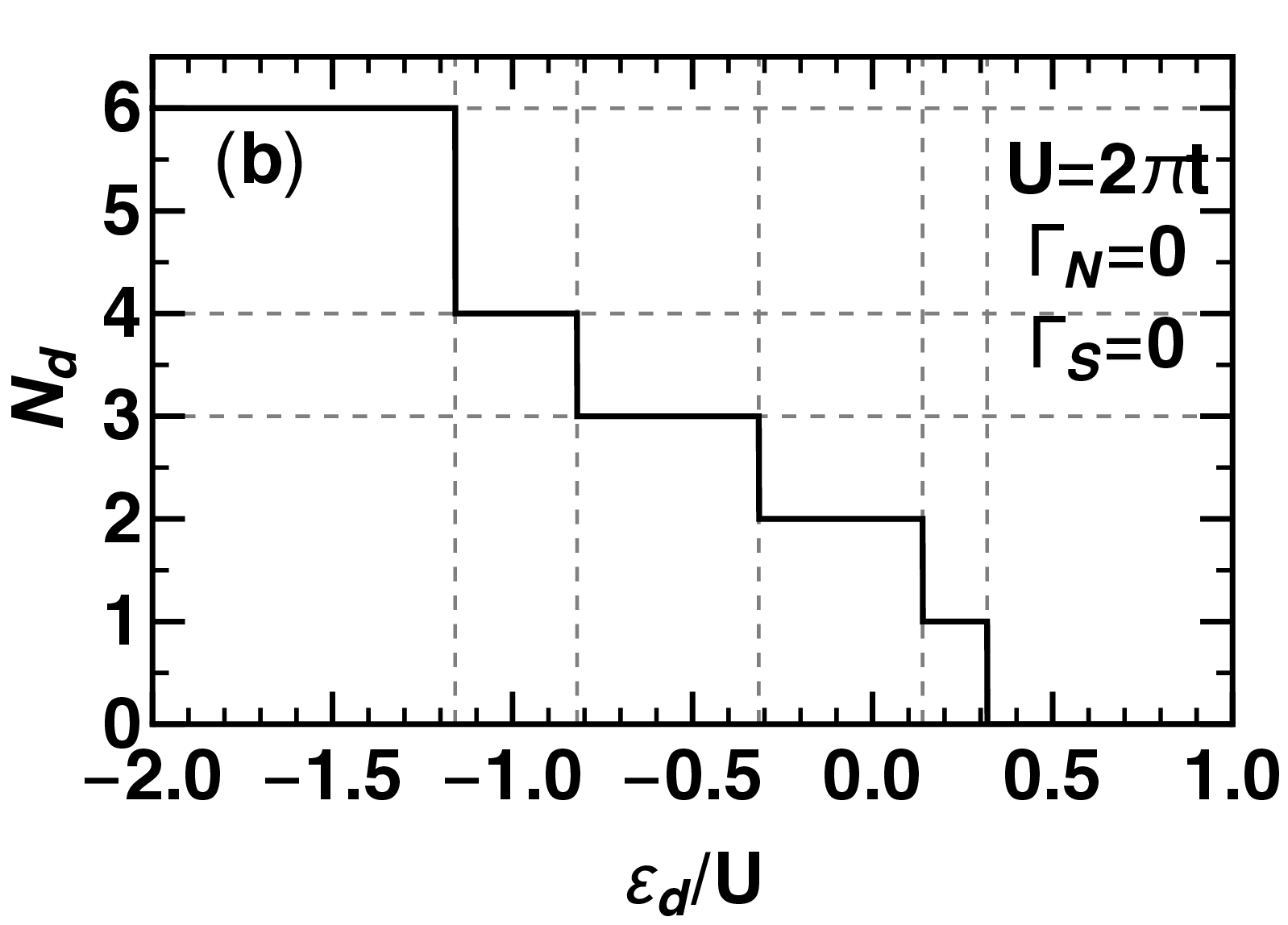}&
		\includegraphics[width=0.38\linewidth]{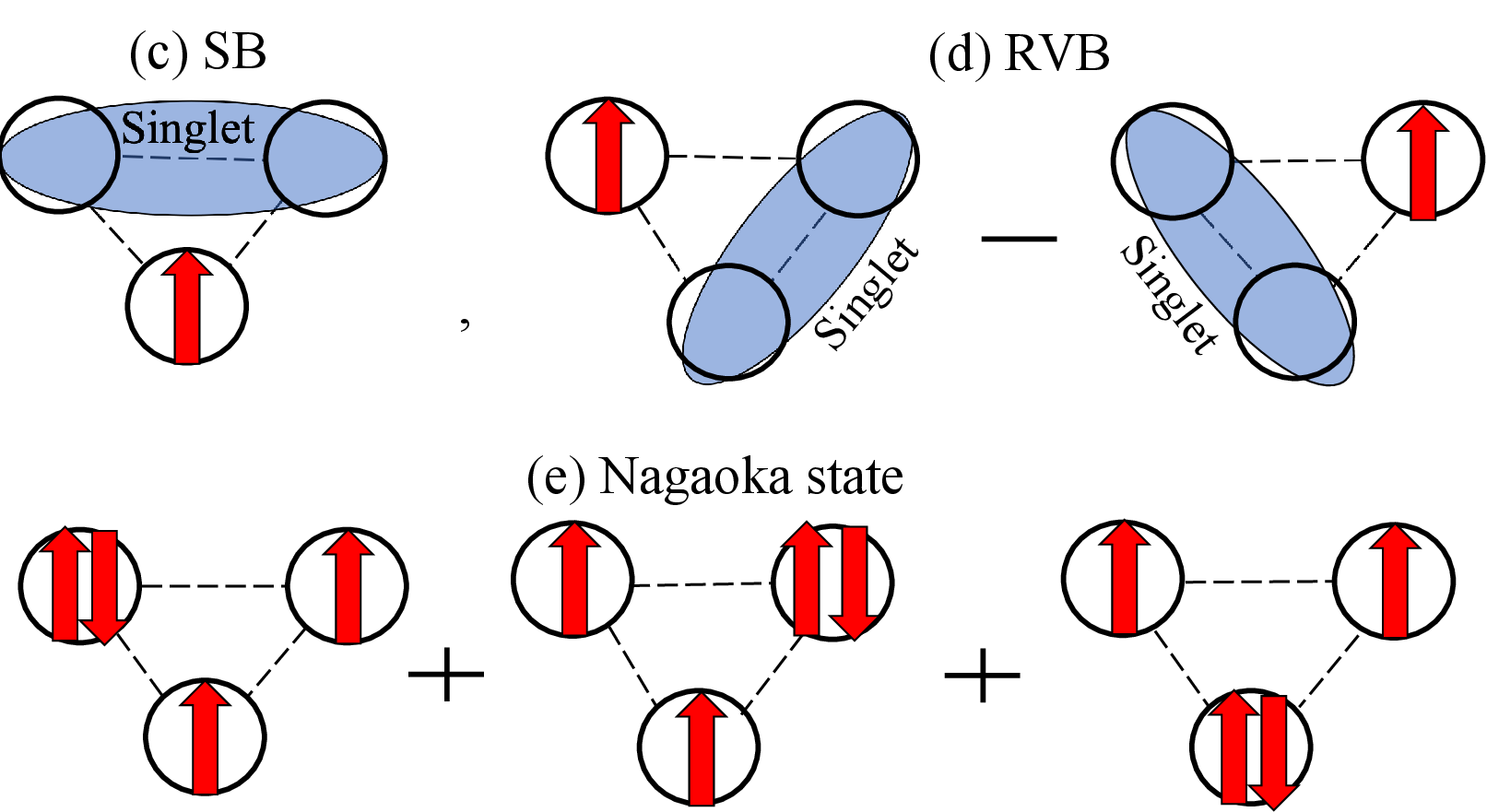}
	\end{tabular}
\caption{
(a): TTQD connected to  one normal (N) and two SC leads.
(b): Occupation number $N_d$ of the isolated TTQD cluster 
plotted vs  $\varepsilon_d$ for $U=2\pi t$. 
Electron configurations in the  
(c) Singlet Bond, (d) Resonance Valence Bond,  and 
 (e) $S=1$ Nagaoka states,
defined in Eqs.\ \eqref{eq:SB_RVB_states}-\eqref{eq:Nagaoka_satate},
are illustrated  for $U \gg t$. 
}
\label{fig:QDsystem}
\end{figure}

\section{Formulation}

We consider the TTQD connected to one normal and two SC leads, 
shown schematically in Fig.\ref{fig:QDsystem}(a) and 
can be described by the Hamiltonian
of the form 
$\ H  = 
H_{\text{dot}} + H_{\text{N}} + H_{\text{TN}} + H_{\text{S}} + H_{\text{TS}}$:
\begin{align}
H_{\text{dot}} =\ &
\sum_{i=L,R,N} \sum_\sigma \, \varepsilon_d^{}\, 
n_{d,i,\sigma}  
\, + \, U \sum_{i=L,R,N} n_{d,i,\uparrow}n_{d,i,\downarrow} 
\, - \, t \sum_{<i,j>} \left( d^\dagger_{i,\sigma}d^{}_{j,\sigma} 
+d^\dagger_{j,\sigma}d_{i,\sigma}^{} \right) \ ,
	\\
	  H_{\text{N}}^{} =& \ \sum_{\sigma} 
	\int_{-D}^{D}  \! d\varepsilon \,\varepsilon\,
	c^\dagger_{\varepsilon,\sigma}c^{ }_{\varepsilon,\sigma} \ , \qquad \qquad 
	H_{\text{TN}}^{} =  v_{N}
	\sum_{\sigma} 
	\int_{-D}^{D}  \! d\varepsilon \,\sqrt{\rho_c^{}}\,
	\Bigl( c^\dagger_{\varepsilon,\sigma}d^{}_{N,\sigma} 
	+ \mathrm{H.c.}
	\Bigr) \ ,
	\\
	H_{\text{S}} =& \   
	\sum_{\alpha=L,R}
	\left[\,
	\sum_{\sigma} 
	\int_{-D_S^{}}^{D_S^{}}  \!\! d\varepsilon \,\varepsilon\,
	s^\dagger_{\varepsilon, \alpha, \sigma} s^{}_{\varepsilon, \alpha, \sigma} 
	+ \! 
	\int_{-D_S^{}}^{D_S^{}}  \!\! d\varepsilon 
	\left( \Delta_{S,\alpha} \,s^\dagger_{\varepsilon, \alpha, \uparrow} 
	s^\dagger_{\varepsilon, \alpha, \downarrow} 
	+  \mathrm{H.c.} \right)
	\,\right],
	\\
	H_{\text{TS}} = & \  
	v_{\text{S}}^{}
	\sum_{\alpha=L,R}  \sum_{\sigma} \! 
	\int_{-D_S^{}}^{D_S^{}}  \!\!  d\varepsilon \,\sqrt{\rho_S^{}} \, 
	\Bigl( s^\dagger_{\varepsilon, \alpha, \sigma}d^{}_{\alpha, \sigma}
	+ \mathrm{H.c.} 
	\Bigr) \, ,
\qquad \qquad  \rho_S^{} \equiv \frac{1}{2D_S^{}}\,. 
\end{align}
Here, 
$d^\dagger_{i,\sigma}$ ($d^{}_{i,\sigma}$) is the creation (annihilation) operator 
for an electron with energy 
$\varepsilon_d^{}$
and spin $\sigma$ in the quantum dot $i$ ($= L,R,N$), 
$n_{d,i,\sigma} \equiv d^\dagger_{i,\sigma} d^{}_{i,\sigma}$ 
is the local number operator, 
 $U$ is the Coulomb interaction,
and $t$ ($>0$) is the hopping matrix element between the dots.
$c^\dagger_{\varepsilon,\sigma}$  and $c^{}_{\varepsilon,\sigma}$ are 
 the operators for conduction electrons in the normal lead, 
defined such that  
$\left \{ c^{}_{\varepsilon,\sigma} \ , c^{\dagger}_{\varepsilon ', \sigma '} \right \} 
= \delta_{\sigma, \sigma '} \,\delta(\varepsilon - \varepsilon')$, 
with  $\rho_c \equiv 1/(2D)$ the density of states  and $D$ the half-band width.   
$v_N$ is the tunneling matrix element between the dot and the normal lead 
and it determines the resonant width  $\Gamma_N \equiv \pi \rho_c v_N^2$. 
The operators 
 $s^\dagger_{\varepsilon,\alpha,\sigma}$ and $s^{}_{\varepsilon,\alpha,\sigma}$ 
describe  electrons in the SC leads on $\alpha= L$, $R$, with 
$\Delta_{S,\alpha} \equiv |\Delta_{S,\alpha}|\, e^{i\theta_\alpha}$ the SC 
energy gap.

This model takes a simplified form 
in the large gap limit  $|\Delta_{S,\alpha}| \to \infty$, 
where 
the SC proximity effects can be described by 
the pair potentials penetrating into the adjacent dots 
\cite{Tanaka2007}, 
\begin{equation}
 H_{\text{S}} + H_{\text{TS}} 
\,\xrightarrow{\,|\Delta_{S,\alpha}| \to \infty\,} \, \sum_{\alpha=L,R}
 \Gamma_S^{} \left( \, 
e^{i\theta_\alpha} d^\dagger_{\alpha,\uparrow} d^\dagger_{\alpha,\downarrow}
+ e^{-i\theta_\alpha} d^{}_{\alpha,\downarrow}d^{}_{\alpha,\uparrow}
 \right) \ ,
\qquad \qquad 
\Gamma_S^{} \equiv \pi  \rho_S^{} v^2_S
\,.
\label{eq:pair_correlation_TTQD}
\end{equation}
In this report, we investigate the case where 
the Josephson effect is absent, i.e., $\theta_L =\theta_R=0$.

\section{Various kinds of local moments induced in the isolated TTQD}

We discuss here some notable properties that 
the TTQD already has 
in the isolated limit  $\Gamma_N = \Gamma_S =0$.
Figure \ref{fig:QDsystem}(b) shows the occupation number 
$N_d \equiv \sum_{i} N_{d,i}$, defined by 
 $N_{d,i} \equiv \sum_\sigma \left \langle  n_{d,i,\sigma} \right\rangle$ 
at zero temperature $T=0$, 
as a function of $\varepsilon_d$ for $U=2\pi t$.  
In the plateau region of
the half-filled case $N_d = 3$, 
the ground state has a four-fold degeneracy 
which can be classified into two parts, 
referred to as the singlet bond (SB)  
and the resonance valence bond (RVB) states for $U \gg t$   
\cite{Mitchell2009,Numata2009},
\begin{align}
 \left| \Phi_{\text{SB}};\sigma\right\rangle \, \equiv  & \  
  \frac{1}{\sqrt{2}} \ d^\dagger_{N,\sigma}
\left( d^\dagger_{L,\uparrow}d^\dagger_{R,\downarrow}
 - d^\dagger_{L,\downarrow}d^\dagger_{R,\uparrow}\right) 
\left|\, 0\right\rangle \ , \label{eq:SB_RVB_states}\\
  \left| \Phi_{\text{RVB}};\sigma\right\rangle \, \equiv & \ 
  \frac{1}{\sqrt{6}} \left[ \,
  d^\dagger_{L,\sigma}
\left( d^\dagger_{R,\uparrow}d^\dagger_{N,\downarrow} 
- d^\dagger_{R,\downarrow}d^\dagger_{N,\uparrow}\right) 
  -  d^\dagger_{R,\sigma}
\left( d^\dagger_{L,\uparrow} d^\dagger_{N,\downarrow} 
- d^\dagger_{L,\downarrow}d^\dagger_{N,\uparrow}  \right) \, \right] 
\left| \,0\right\rangle \ .
\end{align}
In the SB state,  a single unpaired spin $\sigma$ 
is localized in the dot adjacent to the normal lead. 
Therefore, this local moment can easily be 
screened by conduction electrons to form the Kondo singlet 
when the tunnel coupling $H_{\text{TN}}$ is switched on.
In contrast,
in the RVB state  the unpaired spin situates in the dot,  $\alpha=L$ or $R$,   
away from the normal lead as illustrated in Fig.\ref{fig:QDsystem}(d).
For this reason, the local moment in the RVB state remains unscreened 
for small  tunnel couplings  $\Gamma_N^{}$, 
as shown in the next section.

In the plateau region of $N_d = 4$ next to the half-filled one,    
the ground state becomes a spin triplet  for $U> 0$ and $t>0$. 
This is
caused by the Nagaoka ferromagnetic mechanism along the closed loop,   
which is illustrated in Fig.\ \ref{fig:QDsystem}(e) 
for the spin $S_z =+1$ component,
\begin{align}
 	  \left| \Phi_{\text{Nagaoka}};  \,S_z =+1 \right\rangle \, 
 	  \equiv \, \frac{1}{\sqrt{3}} 
\left( d^\dagger_{R,\downarrow} + d^\dagger_{L,\downarrow} 
+ d^\dagger_{N,\downarrow}\right)\,  d^\dagger_{L,\uparrow} d^\dagger_{R,\uparrow}  d^\dagger_{N,\uparrow} \left| \,0\right\rangle \ .
\label{eq:Nagaoka_satate}
\end{align}
This high-spin $S=1$ moment 
can be fully screened via a two-stage processes  
if two conducting channels are coupled 
\cite{Numata2009}.
However, in the configuration shown in Fig.\  \ref{fig:QDsystem}(a), 
 only one-half of the moment 
can be screened by conduction electrons from the single normal lead. 
Furthermore, 
 the leads attached to the TTQD  break the $C_{3v}$ symmetry 
and lift the four-fold degeneracy of the ground state at the plateau of $N_d = 3$. 
We show in the following that 
at small $\Gamma_N$ 
a phase transition occurs between the Kondo-screened SB and the unscreened spin-$1/2$
RVB states  
in the region of  $N_d \simeq 3.0$, 
and then the RVB state continuously evolves into 
the under-screened Nagaoka state as $N_d$ approaches $4.0$.

\begin{figure}[t]
	\centering
	\begin{tabular}{cc}
		\includegraphics[width=0.45\linewidth]{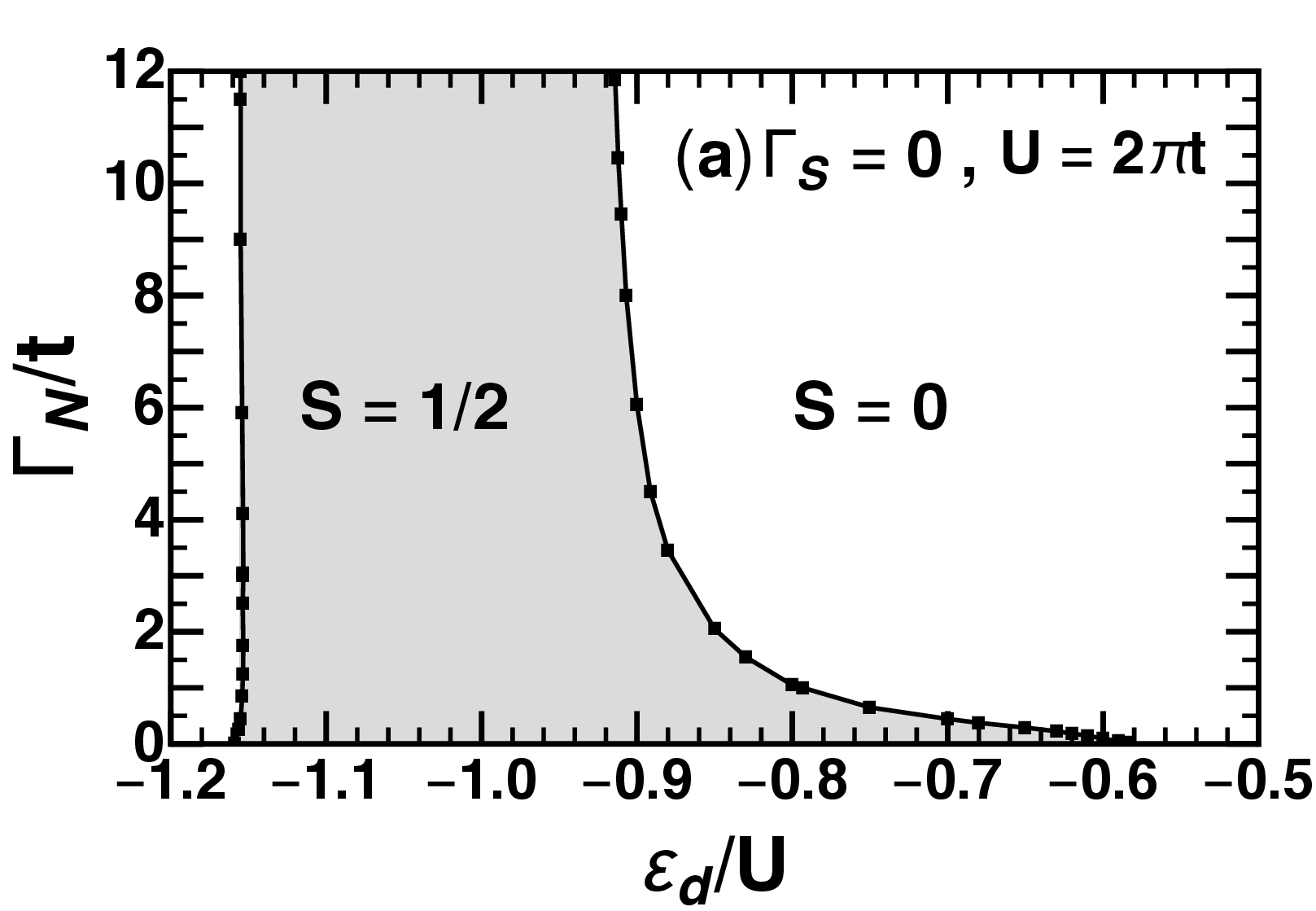}&
		\includegraphics[width=0.45\linewidth]{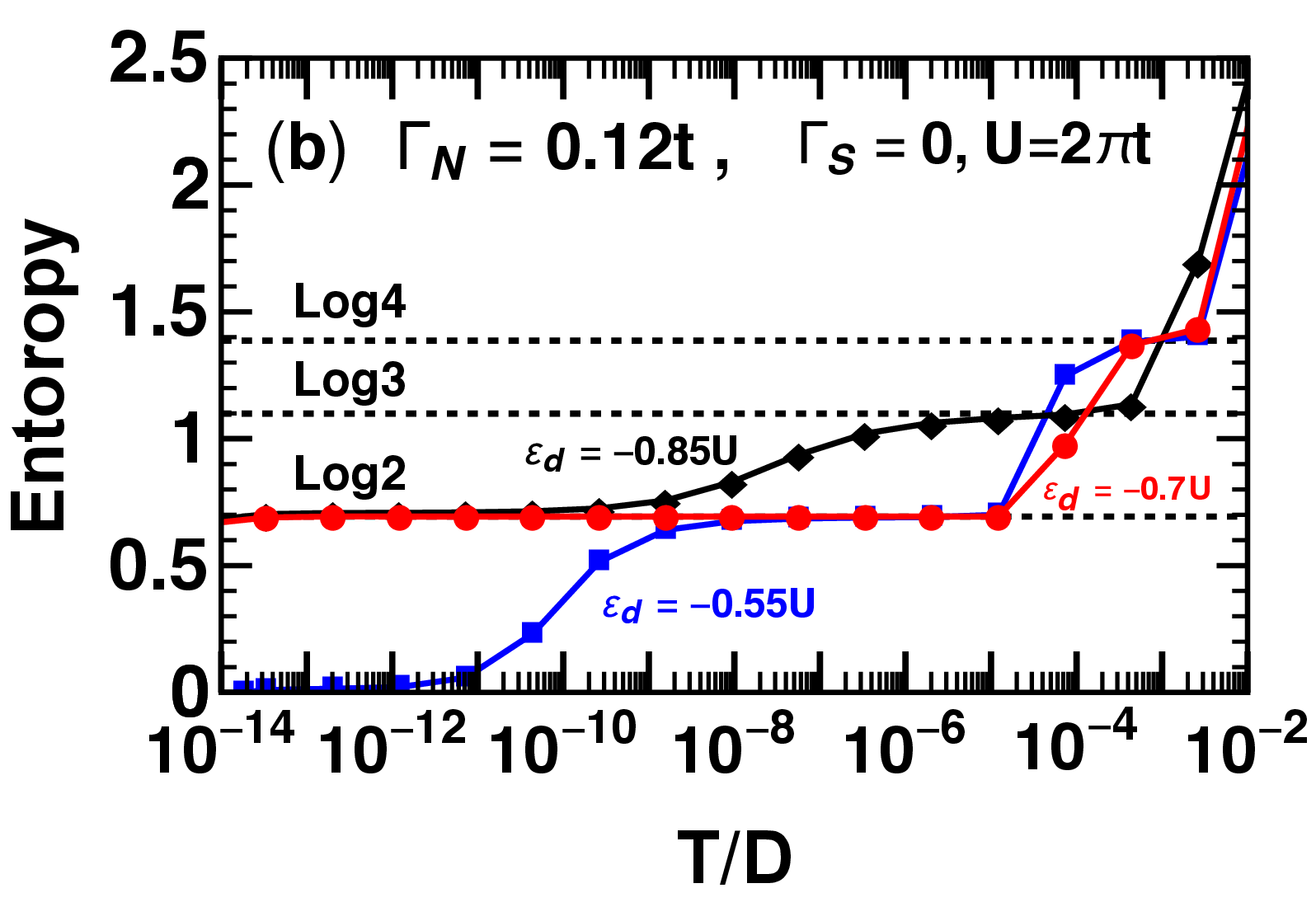}\\
		\includegraphics[width=0.45\linewidth]{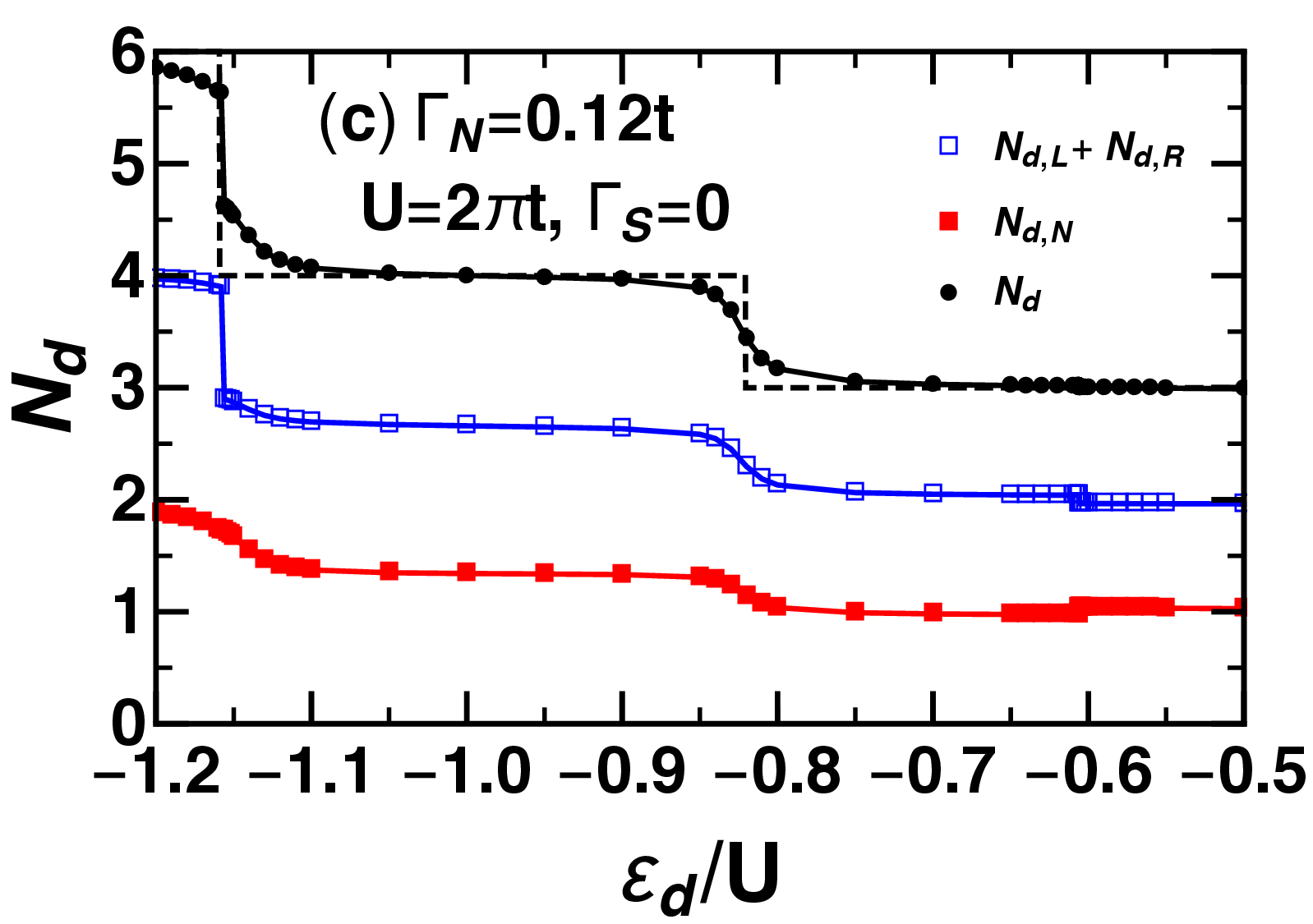}&
		\includegraphics[width=0.45\linewidth]{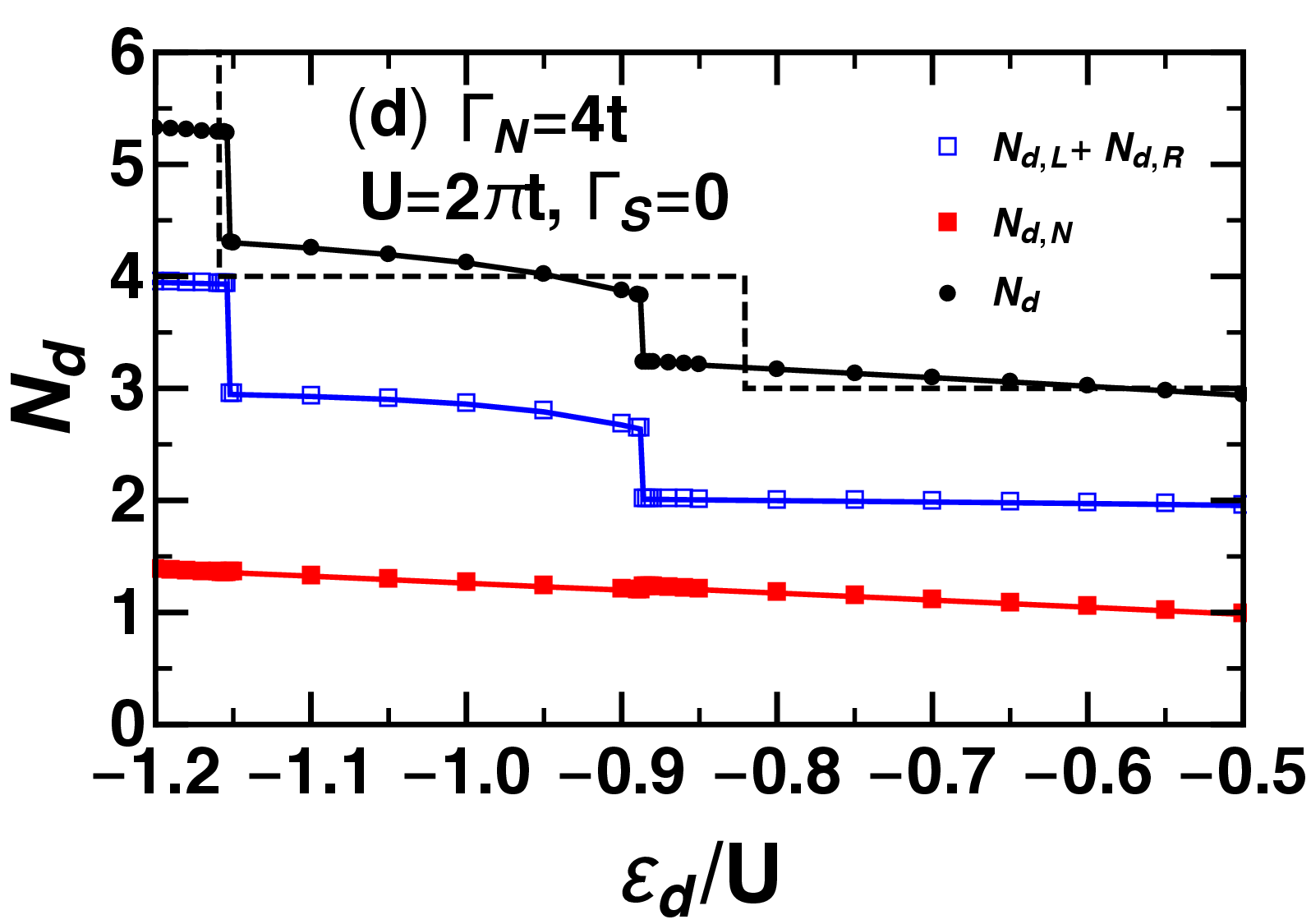}
	\end{tabular}
\caption{
NRG results for TTQD connected to a single normal lead 
($\Gamma_S=0$) for  $U=2\pi t$.
(a): Ground-state phase diagram, 
classified according to the total spin, 
 $S=1/2$ (shaded region) or  $S=0$ (the outside). 
(b): Temperature dependence of the entropy of TTQD 
for $\Gamma_N =0.12t$  at 
 $\varepsilon_d =-0.85U ({\color{black}\blacklozenge}),$
 $-0.70({\color{red}\bullet})$,  
and $-0.55U ({\color{blue}\blacksquare})$. 
The occupation number  $N_{d,N}$ of the dot adjacent to 
the normal lead, that for the other dots $N_{d,L}+N_{d,R}$, 
and the total one $N_{d}$ are plotted vs $\varepsilon_d^{}$ 
for $\Gamma_N = 0.12t$ (c),  and $4t$ (d).
The dash line represents $N_{d}$ for $\Gamma_N=0$.
}
\label{fig:Gapd0}
\end{figure}

\section{Kondo screening of the local moments by a single conducting channel}

In order to clarify how the different kinds 
of local moments 
are screened by conduction electrons,  
we first of all examine the case where
  $\Gamma_S^{}=0$ and only a  single 
normal lead is connected to the TTQD.
This configuration of the TTQD has been studied 
by Michell {\it et al.}  at half-filling $N_d=3$ \cite{Mitchell2009}.
However, here we explore low-energy properties 
over a wide filling range  $3.0 \lesssim N_d \lesssim 6.0$, 
using the numerical renormalization group (NRG) approach \cite{Willson1980}.
We choose the NRG discretization parameter to be $\Lambda =6.0$ 
and retain typically $N_{\text{kept}}=1000$ low-lying excited states.

In Fig.\ \ref{fig:Gapd0}(a),    
the ground-state phase diagram 
is plotted as a function of $\varepsilon_d^{}/U$ and $\Gamma_N/t$ 
for relatively large interaction $U=2 \pi t$. 
The dot-lead coupling  $\Gamma_N$ 
lifts the four-fold degeneracy of the lowest energy state 
of the TTQD cluster near half-filling $N_d^{} \simeq 3.0$. 
The NRG results show that  the ground state 
becomes a doublet in the boot-shaped shaded region   
while it becomes a spin singlet on the outside. 
The spin $S=1/2$ ground state at the tip of the boot,  
which spreads over the region $-0.82 U\lesssim \varepsilon_d^{} \lesssim -0.58U$  
for small 
couplings 
 $\Gamma_N^{} \lesssim 1.5t$, 
can be identified as the RVB state with the free spin-$1/2$ degrees of freedom.
% with the occupation number $N_d^{} \simeq 3.0$. 
Furthermore, the spin $S=0$ ground state on the outside of the tip, 
$-0.58 U\lesssim \varepsilon_d^{} \lesssim -0.32U$, 
can be identified  as the Kondo-screened SB state. 
These identifications can be deduced 
from Figs.\ \ref{fig:Gapd0}(b) and (c). 
% the results shown in 

We see in Fig.\ \ref{fig:Gapd0}(b) that 
the entropy
of
the TTQD calculated 
at the two points, 
($\varepsilon_d=-0.70U$, $\Gamma_N^{}=0.12t$)
and 
($\varepsilon_d=-0.55U$, $\Gamma_N^{}=0.12t$), 
show a similar behavior at high temperatures $T/D \gtrsim 10^{-8}$: 
 both show a plateau of the height $\log 4$ 
reflecting the four-fold degeneracy of the isolated limit  
% due to an approximate four-fold degeneracy  
at temperatures of the order of $T /D \simeq 10^{-3}$, 
and then at  $10^{-8} \lesssim T /D \lesssim 10^{-5}$  
they take another wide plateau of the height $\log 2$ corresponding 
to the spin degrees of freedom of the SB or the RVB state.  
However, at much lower temperatures $T /D \lesssim 10^{-11}$  
the moment vanishes for $\varepsilon_d=-0.55U$, 
whereas the moment remains unscreened 
for $\varepsilon_d=-0.70U$ 
as $\Gamma_N^{}$ is small at this point in the phase diagram.
The entropy at the point ($\varepsilon_d=-0.85U$, $\Gamma_N^{}=0.12t$) 
is also plotted in Fig.\ \ref{fig:Gapd0}(b).  
It takes a plateau of the height  $\log 3$ corresponding to the free  $S=1$ moment 
of Nagaoka state at  $10^{-6} \lesssim  T /D \lesssim  10^{-3}$, 
and then at lower temperatures 
$T /D \lesssim  10^{-9}$ 
it takes another plateau of the height $\log 2$, showing that 
half of this moment is screened by conduction electrons from the normal lead.

Charge distribution of the 
ground state is plotted vs $\varepsilon_d^{}/U$
in Fig.\ \ref{fig:Gapd0}(c) for small dot-lead coupling $\Gamma_N^{}=0.12t$.    
We can see at $\varepsilon_d \simeq -0.6U$, where 
 the phase transition between the singlet 
and doublet ground states takes place, 
the occupation number $N_{d,N}$ of dot on the normal-lead side  
and that of the other two dots  $N_{d,L}+N_{d,R}$ show a small but finite discontinuity. 
At this point the total amount of these occupation numbers 
is almost unchanged  $N_d \simeq 3.0$, 
which indicates the transition occurs between 
the Kondo screened SB and the unscreened RVB states.
As {$\varepsilon_d^{}$} decreases, 
the total occupation number $N_d$ varies from $3.0$ to $4.0$ 
through a gradual step at $\varepsilon_d^{} \simeq -0.82U$. 
This means that the unscreened RVB state emerging at the tip 
of the boot-shaped region can continuously evolve into
the under-screened Nagaoka state spreading widely over the other side.

In contrast,  
in the leg part of the shaded boot-shaped region in Fig.\ \ref{fig:Gapd0}(a), 
 i.e., for large dot-lead couplings $\Gamma_N \gtrsim 1.5t$, 
the total occupation number $N_d$ discontinuously changes 
its value by 
an approximate  amount of $1.0$
at the 
phase boundary along $\varepsilon_d \simeq -0.9U$,   
as demonstrated in Fig.\ \ref{fig:Gapd0}(d) for $\Gamma_N =4t$.
Therefore, the under-screened Nagaoka state spreads
across the leg part of the boot to 
the boundaries on both sides 
$\varepsilon_d \simeq -0.9U$ and $\varepsilon_d \simeq -1.15U$.

\begin{figure}[t]
	\centering
	\begin{tabular}{cc}
		\includegraphics[width=0.45\linewidth]{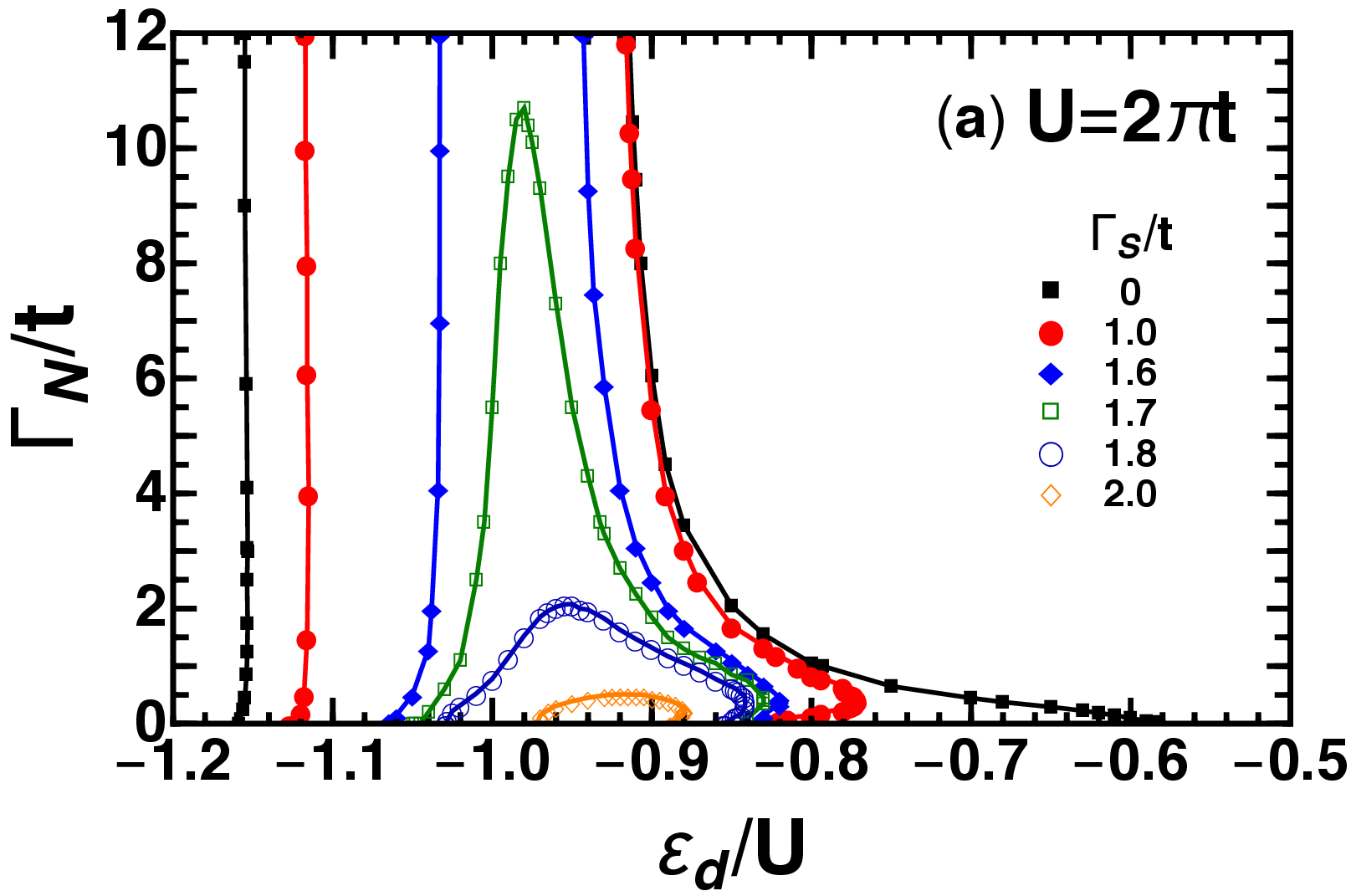}&
		\includegraphics[width=0.45\linewidth]{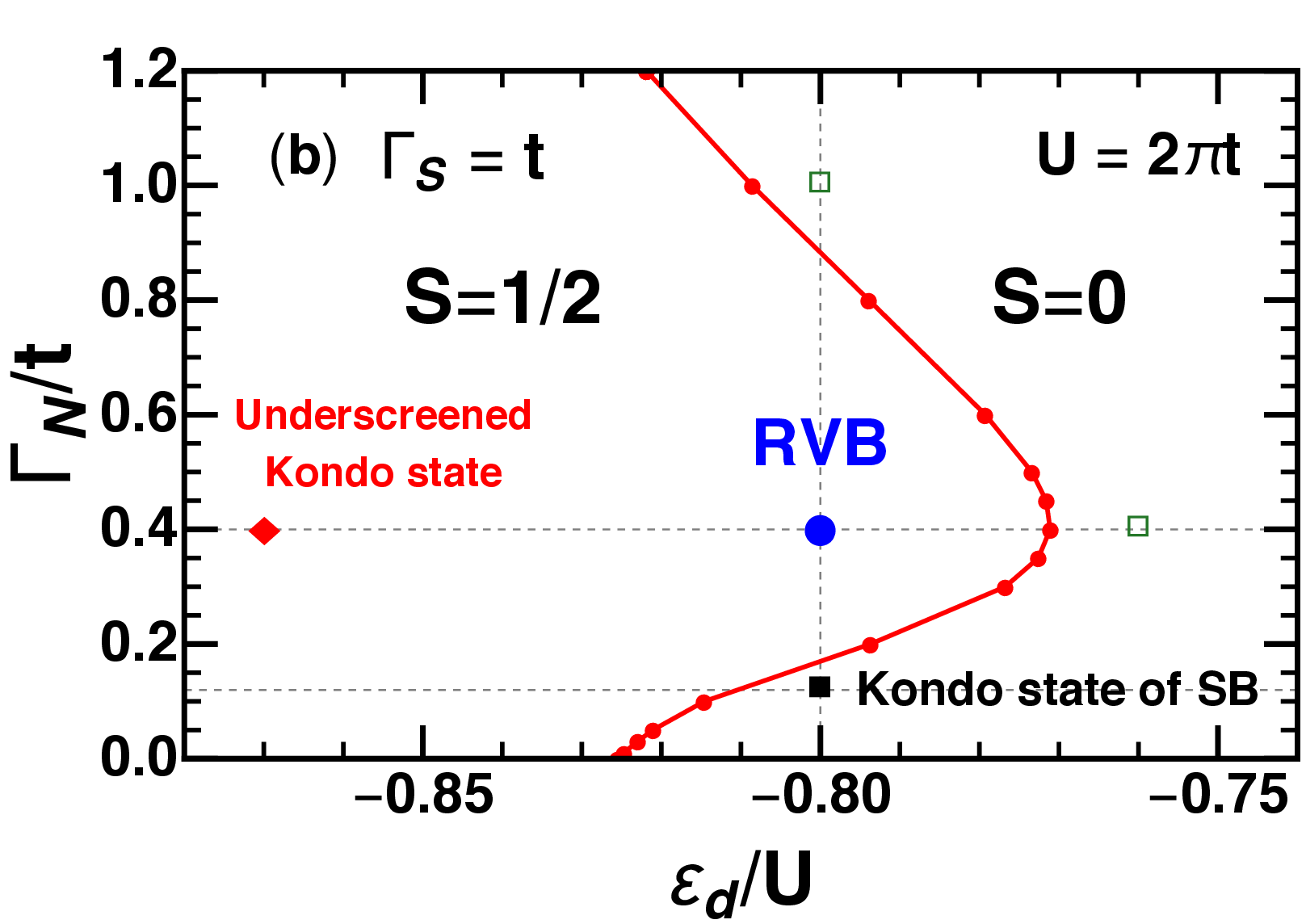}\\
		\includegraphics[width=0.47\linewidth]{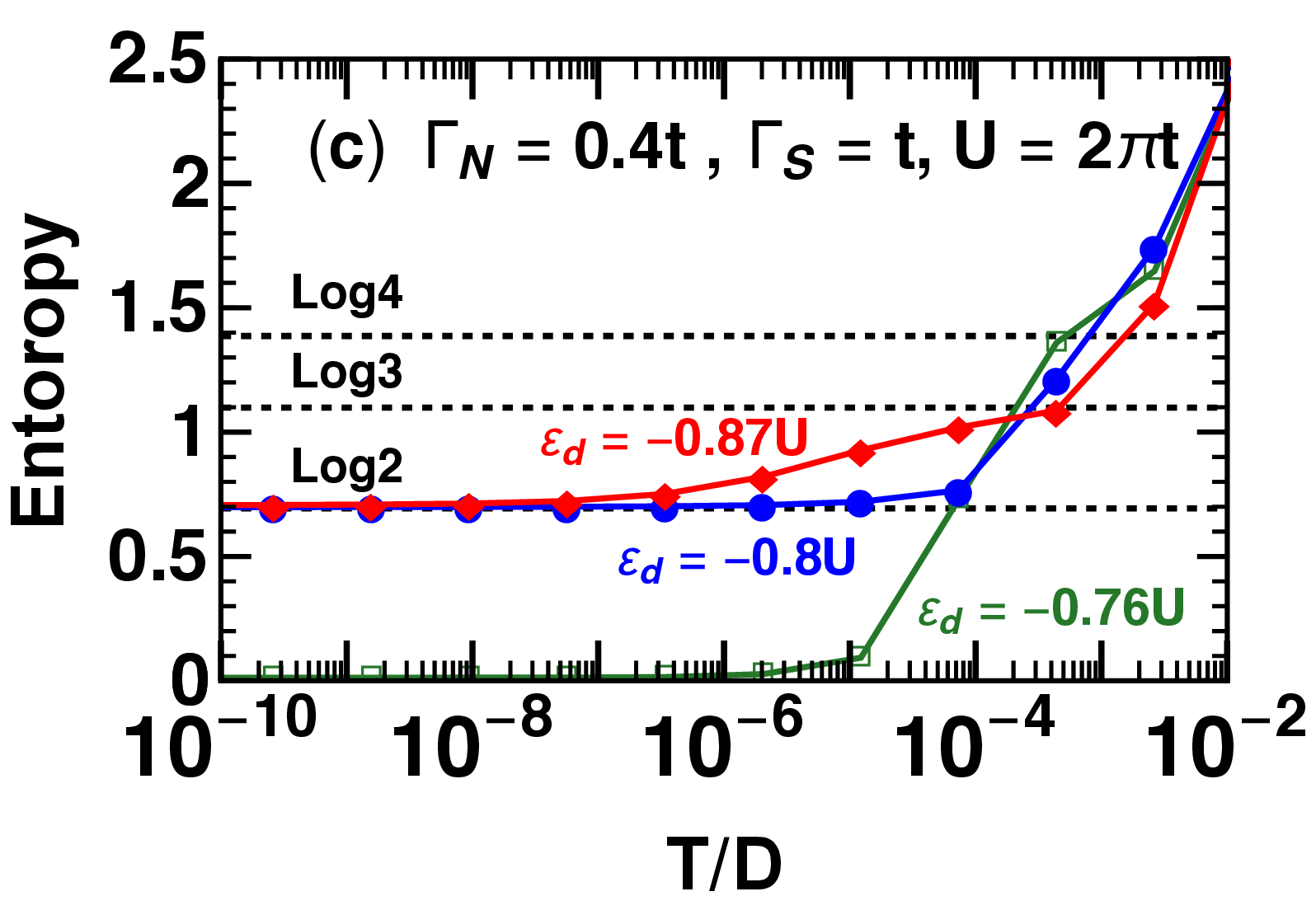}&
		\includegraphics[width=0.47\linewidth]{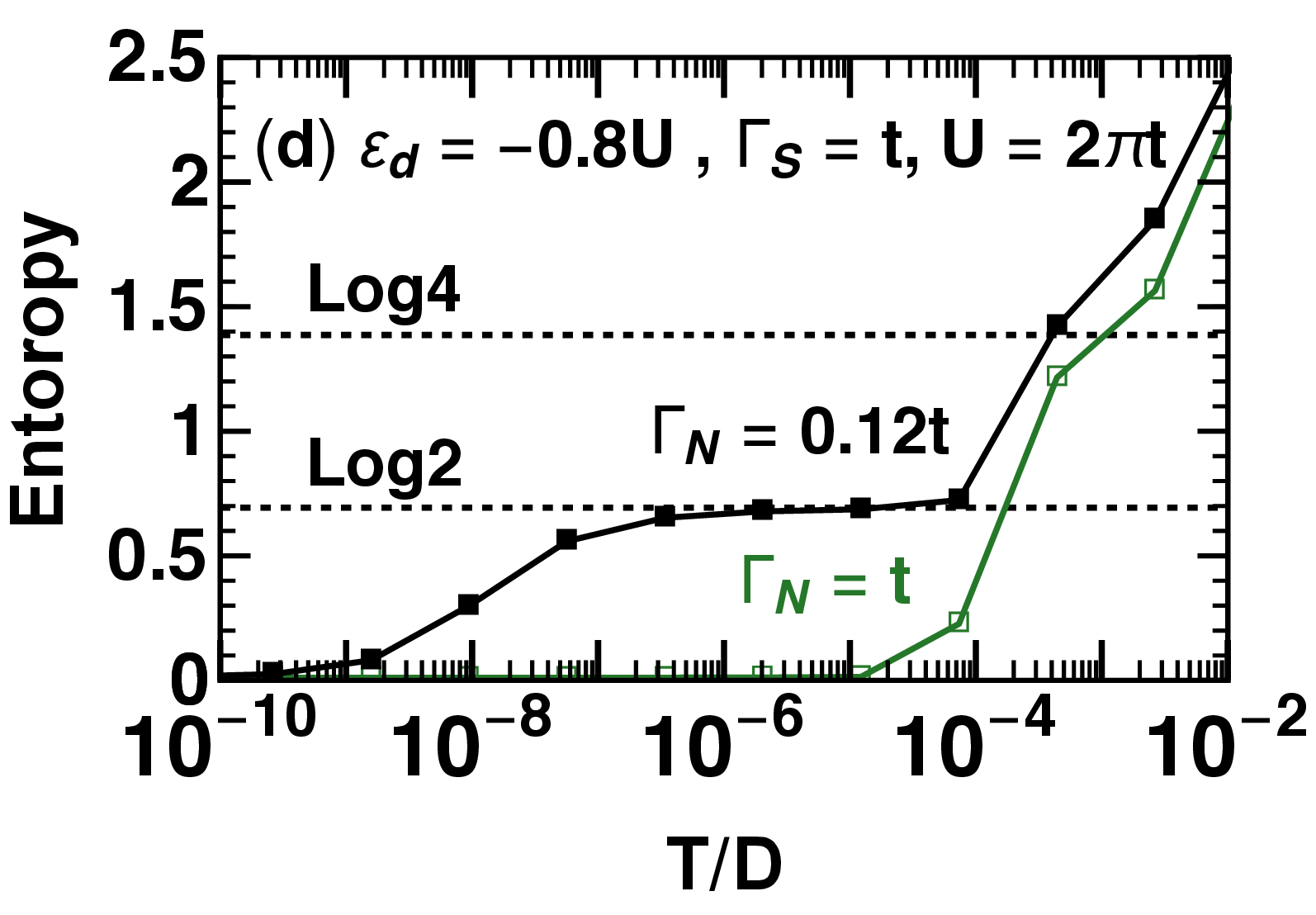}
	\end{tabular}
\caption{
NRG results for TTQD connected to one normal and two SC leads for  $U= 2\pi t$.
(a): Ground-state phase diagram classified according to the total spin, 
 $S=1/2$ (enclosed region) or $S=0$ (the outside), are plotted 
for $\Gamma_S/t=0$, $1.0$, $1.6$, $1.7$, $1.8$, $2.0$.
(b): An enlarged view of a reentrant phase boundary for $\Gamma_S=t$.
 Temperature dependence of the entropy of TTQD  for $\Gamma_S = t$ is plotted 
in (c) at  $\varepsilon_d = -0.76U ({\color[rgb]{0.15, 0.47, 0.17} \square})$, 
$-0.80U ({\color{blue} \bullet})$, 
$-0.87U ({\color{red} \blacklozenge})$ 
for $\Gamma_N = 0.4t$,  
and in (d) 
at $\varepsilon_d  =-0.80U$  
 for 
 $\Gamma_N  = 0.12t ({\color{black} \blacksquare})$, 
$1.0 t ({\color[rgb]{0.15, 0.47, 0.17} \square})$. 
}
\label{fig:GroundstateGapd}
\end{figure}

\section{Ground state of the TTQD connected to one normal and two SC leads}

We next consider the configuration with two additional SC leads 
as illustrated  in Fig.\ \ref{fig:QDsystem}(a). 
The ground-state phase diagram is shown in
Fig.\ \ref{fig:GroundstateGapd}(a) 
as a function of $\varepsilon_d/U$ and $\Gamma_N/t$ for several values of 
$\Gamma_S/t$,
choosing the same value $U=2\pi t$ for the interaction as in the above. 
We can see that the region of the $S=1/2$ ground state shrinks 
as couplings to the SC leads $\Gamma_S$, 
 defined in Eq.\ \eqref{eq:pair_correlation_TTQD}, 
 increases. 
In particular for $\Gamma_S/t =2.0$, 
the under-screened $S=1/2$ region around  
$\varepsilon_d \simeq -0.93U$ becomes very narrow 
and is enclosed by the boundary line 
as Cooper pairs penetrating from the SC leads 
 dominate the other magnetic correlations.
We can also see that a reentrant transition 
occurs  for small $\Gamma_N$ across the $S=1/2$ region  
at $-0.9U \lesssim \varepsilon_{d} \lesssim -0.6U$.
This is caused by the fact that an infinitesimal $\Gamma_S$ 
lifts the four-fold degeneracy of the lowest energy state in the half-filled case $N_d=3$  
and makes the SB state the ground state 
for very small normal tunnel couplings $\Gamma_N\ll t$.
The singlet region which spreads below the doublet region, 
for instance the one emerges at 
$\Gamma_N/t \lesssim 0.4$ in Fig.\ \ref{fig:GroundstateGapd}(b), 
can be identified as the Kondo-screened SB state. 
The other side of the boundary can be identified as 
 the unscreened RVB state for which effects of  
 $\Gamma_N$ dominates that of $\Gamma_S$. 
These identifications
can be verified from the behaviors of the entropy 
of the TTQD, 
calculated for $\Gamma_S=1.0t$  
at five different points on the dashed-vertical 
and dashed-horizontal lines in the phase diagram    
Fig.\ \ref{fig:GroundstateGapd}(b).

The entropy of the TTQD obtained at the three points on 
the horizontal-dashed line for $\Gamma_N =0.4t$ 
are plotted in Fig.\ \ref{fig:GroundstateGapd}(c).  
The entropy 
at the point  $\varepsilon_d = -0.87U$, 
where $N_d \simeq 4.0$,  shows a plateau of the height $\log 3$ 
at high temperatures $10^{-4} \lesssim T /D \lesssim  10^{-3}$, 
and then at low temperatures it converges to $\log 2$, 
which indicates that the ground state at this point is the under-screened Nagaoka state. 
In contrast,  at the point of $\varepsilon_d = -0.80U$ where $N_d \simeq 3.0$,  
the entropy directly approaches the value for the free $S=1/2$ moment as $T$ decreases 
without showing a step at $\log 3$, 
and thus the ground state at this point can be identified as the unscreened RVB state.
We can  also see that 
the entropy at the point  ($\varepsilon_d/U = -0.80$, $\Gamma_N/t = 0.12$) 
 plotted in Fig. \ref{fig:GroundstateGapd}(d) 
clearly shows the behavior of the fully-screened SB state:
free $S=1/2$ moment emerging at intermediate temperatures 
 $10^{-6} \lesssim  T /D \lesssim  10^{-4}$ vanishes at low temperatures.
The  entropy calculated at the other two points, 
 ($\varepsilon_d/U = -0.76$,  $\Gamma_N/t = 0.4$) and 
 ($\varepsilon_d/U = -0.80$,  $\Gamma_N/t = 1.0$),    
exhibit 
a similar behavior,  showing 
a tendency to 
take  a plateau of the height $\log 4$ at high temperatures, 
and then rapidly decrease 
with decreasing temperatures, 
without showing any other structures at intermediate temperatures.

We can also see in Fig.\ \ref{fig:GroundstateGapd}(a) 
that the reentrant behavior is suppressed 
as tunnel coupling to the SC lead $\Gamma_S$ increases.  
The unpaired spin of the RVB state, illustrated 
in Fig.\ \ref{fig:QDsystem}(d),  can be 
replaced by local Cooper pairs penetrating 
into the adjacent dots  from the SC leads for large $\Gamma_S$, 
and it makes the ground state a singlet near the reentrant region.

\section{Summary}

We have studied the quantum phase transition and crossover occurring 
in the TTQD over a wide range of electron fillings.   
In the case where only a single normal lead is connected,  
the unscreened RVB state and the under-screened Nagaoka state 
merge 
into 
the doublet phase,
which appears as a boot-shaped region in the phase diagram. 
The RVB state spreading over the tip of the boot 
can continuously evolve into the under-screened Nagaoka state 
as $\varepsilon_d$ decreases.
When two additional SC leads are connected, 
a reentrant transition takes place 
around the tip part of the doublet phase as
the SC proximity makes the Kondo-screened SB state 
the ground state for small $\Gamma_N \ll t$. 
The magnetic doublet phase shrinks as Cooper pairs 
penetrating into the TTQD 
increase with  $\Gamma_S$.

\section*{Acknowledgments}

This work was supported by JSPS KAKENHI Grant No. JP18K03495, JST Moonshot R\&D - MILLENNIA Program Grant No. JPMJMS2061, 
and the Sasakawa Scientific Research Grant from the Japan Science Society No.2021-2009.

\end{document}